\documentclass[12pt]{article}

\usepackage{amssymb}
\usepackage{amsmath}
\usepackage{array} 
\usepackage{epsfig}
\usepackage{graphics}

\begin{document}                              

\begin{center}
            
 {\Large \bf                       
\vskip 7cm                  
\mbox{LHC as $\pi p$  and $\pi\pi$ Collider.}
}
\vskip 1cm

\mbox{V.A.~Petrov, R.A.~Ryutin and A.E.~Sobol}

\mbox{{\small Institute for High Energy Physics}}

\mbox{{\small{\it 142 281} Protvino, Russia}}

 \vskip 1.75cm
{\bf
\mbox{Abstract}}
  \vskip 0.3cm

\newlength{\qqq}
\settowidth{\qqq}{In the framework of the operator product  expansion, the quark mass dependence of}
\hfill
\noindent
\begin{minipage}{\qqq}
We propose an experiment at the LHC with leading neutron production.The latter can be used to extract from it the total $\pi^+ p$ cross-sections.
With two leading neutrons we can get access to the total $\pi^+\pi^+$ cross-sections.  In this note we give some estimates and discuss related problems and prospects.

\end{minipage}
\end{center}


\begin{center}
\vskip 0.5cm
{\bf
\mbox{Keywords}}
\vskip 0.3cm

\settowidth{\qqq}{In the framework of the operator product  expansion, the quark mass dependence of}
\hfill
\noindent
\begin{minipage}{\qqq}
Leading Neutron Spectra -- Total cross-section -- Absorption -- Regge-eikonal model
\end{minipage}

\end{center}

\setcounter{page}{1}
\newpage

\section{Introduction}

 LHC opens new possibilities for diffractive physics, especially in measurements of the total and elastic $pp$ cross-sections. This will allow to discriminate among many models of high-energy diffractive scattering. However it is not enough and for more distinctive separation of viable models we badly need the information on the high-energy cross-sections of other initial states. There are also some quite general considerations,e.g., a universal high-energy behavior of any total cross-section, independently of the initial state. Unfortunately, other processes are left far behind $pp$ and $\bar{p} p$. For instance, the total cross-section of $\pi^+ p$ interaction is known only up to 25 GeV.
 At present no plans exist to get high-energy secondary beams to fill this gap. Nonetheless we could –- due to an old idea of Goebel and Chew-Low~\cite{chewlow},\cite{goebel} –-
 try to use indirect methods. Earlier there were already attempts to do that. For example in Refs.~\cite{pipiextract},\cite{pipiextract2} total and elastic $\pi\pi$ cross-sections 
 were extracted in the energy domain 1.5 –- 4.0 GeV from the cross-sections of exclusive processes with charge exchange. More recent extraction of the $\pi p$ scattering from the data on $\gamma +p\to\pi^+ +\pi^- +p$ was undertaken in Ref.~\cite{pipHERA1} with a (model dependent~\cite{pipHERA2},\cite{pipHERA3}) result: $\sigma_{\pi p}(50\; {\rm GeV})=31\pm 2({\rm stat.})\pm 3({\rm syst.})\; {\rm mb}$. 
 
 Certainly, at LHC it would be more difficult to measure exclusive channels but, instead,  inclusive spectra of fast leading neutrons seem to give an excellent occasion to get pion cross-sections at unimaginable energies 1-5 TeV in the c.m.s.
 
  The process of leading neutron production has been studied at several experiments in
 photon-hadron~\cite{HERA1}-\cite{HERA5} and hadron-hadron~\cite{hadr1}-\cite{hadr5}
 colliders. 
  In this paper we consider processes of the type $p+p\to n+X$ and $p+p\to n+X+n$. Recently some calculations were made in~\cite{KMRn1}-\cite{workn2}. In these works authors paid attention basically to the photon-proton reaction, while for hadron collisions the situation was estimated to be not so clear (see~\cite{workn1},\cite{workn2}).
 
  The leading neutron production is dominated by $\pi$ exchange~\cite{KMRn1}-\cite{workn2} and we have a chance to extract total $\pi^+ p$ and $\pi^+\pi^+$ cross-sections. This is a good motivation
 for an experimental study.
 
   Since the energy becomes large, we have to take into account effects of soft rescattering which can be calculated as corrections to the Born approximation. In the calculations of such absorptive effects we use Regge-eikonal approach~\cite{3Pomerons}.
   In the first part of the paper we present our calculations for differential cross-sections, while in the last section the Monte-Carlo simulation and experimetal possibilities are considered with ZDC~\cite{ZDC} of CMS as a key element. 

\section{Kinematics}

\begin{center}
\begin{figure}[t!]
\hskip  3cm \vbox to 9cm 
{\hbox to 9cm{\epsfxsize=9cm\epsfysize=9cm\epsffile{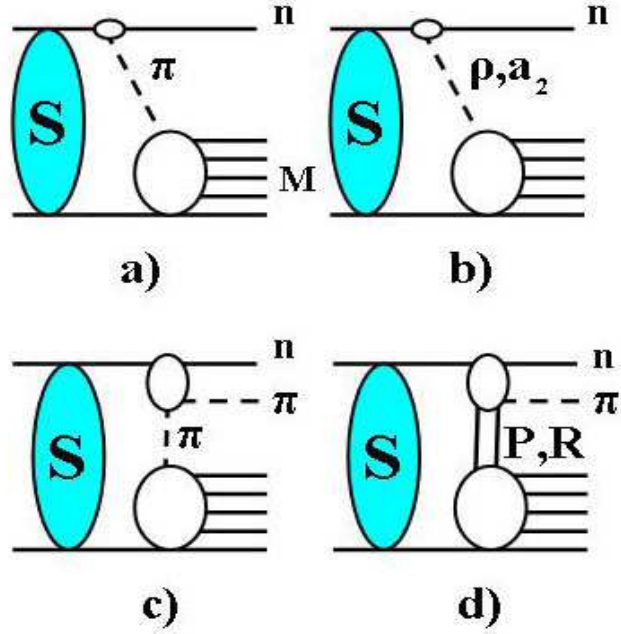}}}
\caption{\label{fig:1}Diagrams for the signal and background processes in $pp$ collisions. a)~process with a single pion exchange (S$\pi$E) $p+p\to n+X$, $M$ is the mass of the system X; b)~process with other reggeon exchanges; c)~double dissociation process with a pion exchange; d)~double dissociation with Pomeron and reggeon exchanges. $S$ represents soft rescattering corrections.}
\end{figure} 
\end{center}

 The diagram of the process $p(p_1)+p(p_2)\to n(p_n)+X(p_X)$ is presented in the Fig.~\ref{fig:1}a. In the center-of-mass frame momenta of particles can be represented as follows (arrows denote transverse momenta):
 \begin{equation}
 \label{kin1a}p_1=\left(\frac{\sqrt{s}}{2},\frac{\sqrt{s}}{2}\beta ,\vec{0}\right),\; p_2=\left(\frac{\sqrt{s}}{2},-\frac{\sqrt{s}}{2}\beta ,\vec{0}\right),
\end{equation}
\begin{eqnarray}
&& \label{kin1c}p_{\pi}=\left( \xi\frac{\sqrt{s}}{2}\beta^2+\frac{t+m_p^2-m_n^2}{\sqrt{s}},\xi\frac{\sqrt{s}}{2}\beta ,\vec{q}\right),\\
&& \label{kin1d}p_n=\left((1-\xi\beta^2)\frac{\sqrt{s}}{2}-\frac{t+m_p^2-m_n^2}{\sqrt{s}},(1-\xi)\frac{\sqrt{s}}{2}\beta ,-\vec{q}\right)\\
&& \label{kin1e}p_X^2=M^2,\; \xi=\frac{M^2-m_n^2-2(t+m_p^2-m_n^2)}{s\beta^2}\simeq\frac{M^2}{s},\\
&& \label{kin1f}-t=\frac{\vec{q}^{\;2}+\xi^2\beta^2m_p^2+(m_n^2-m_p^2)\left( \xi\beta^2-\frac{m_n^2-m_p^2}{s}\right)}{1-\xi\beta^2+\frac{2(m_n^2-m_p^2)}{s}}
\simeq\frac{\vec{q}^{\; 2}+\xi^2m_p^2}{1-\xi},\\
&& \label{kin1g}\beta=\sqrt{1-\frac{4m_p^2}{s}}.
 \end{eqnarray}
 One of the important questions is the definition of the kinematical region of the process, especially in rapidity $y$ (pseudorapidity $\eta$) (see Section~\ref{section:exp}). If we have several secondaries from $\pi^+ p$ scattering with momenta
 \begin{equation}
\label{kin2}
k_i=\left( \sqrt{m_i^2+\vec{k}_i^2+\xi_i^2\beta^2\frac{s}{4}},-\xi_i\beta\frac{\sqrt{s}}{2},\vec{k}_i   \right) 
 \end{equation}
then 
\begin{eqnarray}
&& \sum\limits_i \xi_i=1-\xi,\; \sum\limits_i\vec{k}_i=\vec{q},\nonumber\\
&& \label{kin3}\sum\limits_i\sqrt{\xi_i^2+\frac{4(m_i^2+\vec{k}_i^2)}{s\beta^2}}=1+\xi\beta^2+\frac{t+m_p^2-m_n^2}{\sqrt{s}}\simeq 1+\xi,
\end{eqnarray}
and
\begin{eqnarray}
&& \label{rapid}y_i\simeq \frac{1}{2}\ln\frac{
\left( \sqrt{\xi_i^2+\frac{4(m_i^2+\vec{k}_i^2)}{s}}-\xi_i\right)^2}{\frac{4(m_i^2+\vec{k}_i^2)}{s}},\\
&& \label{pseudorapid}\eta_i\simeq\frac{1}{2}\ln\frac{ 
\left( \sqrt{\xi_i^2+\frac{4\vec{k}_i^2}{s}}-\xi_i\right)^2}{\frac{4\vec{k}_i^2}{s}}.
\end{eqnarray}
For negative $\xi_i$ we have $y_i\to y_{i,max}$,  $\eta_i\to\infty$ for $\vec{k}_i\to 0$. It means that we have no pseudorapidity gap for low momenta of produced hadrons even if we have the rapidity one 
\begin{equation}
y_{i,max}\simeq\ln\frac{\xi_i\sqrt{s}}{m_i}\le\ln\frac{M^2}{\sqrt{s}m_i}.
\end{equation}
Experimentally, it is difficult situation when we need to cut momenta of secondary particles from below (see Section~\ref{section:exp}). For example, if $y=6$ for pions of energy 30 GeV, then $\eta\simeq 9$.

\section{Calculation of the cross-section. Absorptive effects.}

\begin{center}
\begin{figure}[ht!]
\hskip  1cm \vbox to 7cm 
{\hbox to 12cm{\epsfxsize=12cm\epsfysize=7cm\epsffile{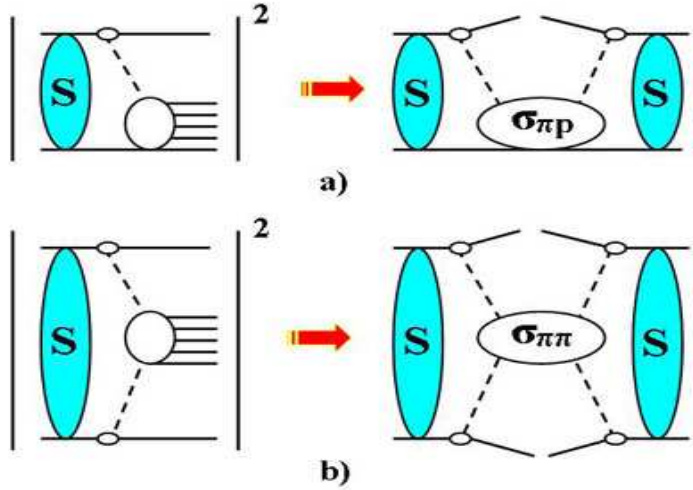}}}
\caption{\label{fig:2} Amplitudes squared and total cross-sections of the processes: a) $p+p\to n+X$ (S$\pi$E), b) $p+p\to n+X+n$ (D$\pi$E). $S$ represents soft rescattering corrections.}
\end{figure} 
\end{center}

 As a Born approximation corresponding to $\pi$ exchange we use familiar triple-Regge formula (see Fig.~\ref{fig:2}), which can be rewritten as follows
\begin{equation}
\label{born}
\frac{d\sigma_0(p+p\to n+X)}{d\xi dt}=\frac{G_{\pi^+pn}^2}{16\pi^2}\frac{-t}{(t-m_{\pi}^2)^2} F^2(t) \xi^{1-2\alpha_{\pi}(t)}\sigma_{\pi^+ p}(\xi s),
\end{equation}
where $\alpha_{\pi}(t)=\alpha^{\prime}_{\pi}(t-m_{\pi}^2)$ is the pion trajectory with slope $\alpha^{\prime}\simeq 0.9$~GeV$^{-2}$, and  $G_{\pi^0pp}^2/(4\pi)=G_{\pi^+pn}^2/(8\pi)=13.75$~\cite{constG}. $\xi=1-x_L$, were $x_L$ is the fraction of initial proton's longitudinal momentum carried by neutron. Form factor $F(t)$ is usually taken in the exponential form
\begin{equation}
\label{formfactor}
F(t)=\exp(bt),
\end{equation}
where, from recent data~\cite{HERA2},\cite{KMRn1c16}, we expect $b\simeq 0.3\; {\rm GeV}^{-2}$.
 We are interested in the kinematical range $0.01$~GeV$^2<|t|<0.5$~GeV$^2$, $\xi<0.4$, where formula~(\ref{born}) dominates~\cite{KMRn1c13},\cite{KMRn1c14}. To make estimates for high energies we can use any adequate parametrization of the total $\pi^+ p$ cross-sections. 
 
\begin{center} 
\begin{figure}[t!]
\hskip  1cm \vbox to 5cm 
{\hbox to 12cm{\epsfxsize=12cm\epsfysize=6cm\epsffile{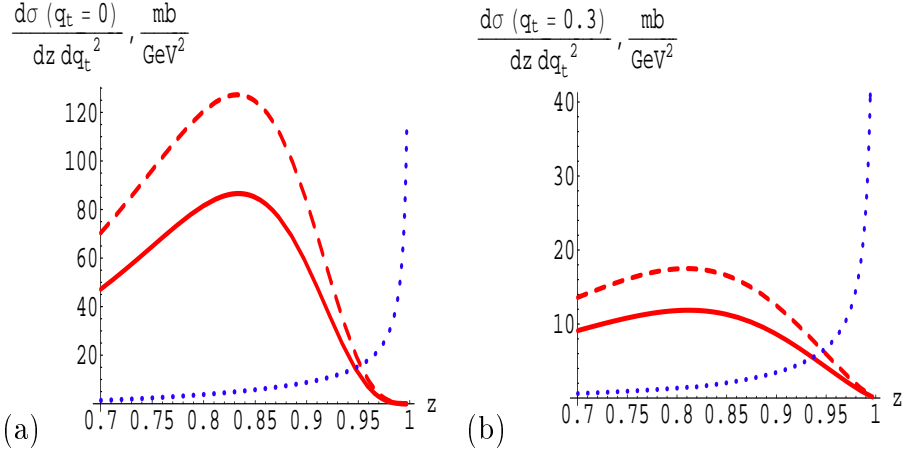}}}
\vskip 0.8cm
\caption{\label{fig:3}Differential cross-sections for the processes $p+p\to n+X$ (parametrization~(\ref{sigland}), solid, and~(\ref{sigcomplete}), dashed) and double dissociation process $p+p\to N^*(\to n+\pi)+X$ (dotted) at different values of transverse momentum transfer versus $z=1-\xi$.}
\end{figure}  
\end{center}

 In addition we should take into account other possible processes with neutron production. We also have to include contributions from $\rho$, $a_2$ exchanges (Fig.~\ref{fig:1}b), and from resonance decays, such as $\Delta$ and $N^*$ in processes shown in Figs.\ref{fig:1}c,d. The calculation of the neutron spectra~\cite{KMRn1c16}-\cite{KMRn1c14} shows that the contribution depicted in Fig.~\ref{fig:1}a dominates, while the contribution corresponding to Fig.~\ref{fig:1}b is about 20\%. Other reggeons give
 also small contribution due to spin effects~\cite{KMRn1c16}. The main
 background can arise from minimum bias events and Fig.~\ref{fig:1}d, which was estimated to be about $0.06\cdot\sigma(p+p\to p+X)$~\cite{ferrari} at low energies. It has, however, inverse missing mass dependence and is suppressed at intermediate $\xi$~(see Fig.~\ref{fig:3}). This fact is used to eliminate the background.
 
 Another important suppression factor arises from absorptive corrections. All possible corrections are discussed in Ref.~\cite{KMRn1}. We estimate only absorption in the initial state, since it gives the main contribution. For this task we use our model with 3 Pomeron trajectories~\cite{3Pomerons}:
\begin{eqnarray}\label{3IPtrajectories}
\alpha_{IP_1}(t)-1&=& (0.0578\pm0.002)+(0.5596\pm0.0078)t \;,\nonumber\\
\alpha_{IP_2}(t)-1&=& (0.1669\pm0.0012)+(0.2733\pm 0.0056)t \;,\nonumber\\
\alpha_{IP_3}(t)-1&=& (0.2032\pm0.0041)+(0.0937\pm0.0029)t \;,
\end{eqnarray}
which are the result of a 20 parameter fit of the total and differential cross-sections in the region $0.01$GeV$^2<|t|<14$GeV$^2$ and $8$GeV$<\sqrt{s}<1800$GeV,   $\chi^2/d.o.f.=2.74$. Although $\chi^2/d.o.f.$ is rather large, the model gives
good predictions for the elastic scattering (especially in the low-t region with $\chi^2/d.o.f.\sim1$). It was also noted in~\cite{3Pomerons}, that this approach may be an artefact of the more general one with Regge cuts or nonlinear Pomeron trajectory. 

Following the procedure described in~\cite{workn1},\cite{workn2}, we
can estimate absorptive corrections. Finally we obtain (an effective factorized form of the following expression~(\ref{Udsigma}) is only used for convenience, there is no factorization):
\begin{equation}
 \label{dsigma0qt}\frac{d\sigma_0(\xi,\vec{q}^{\;2})}{d\xi d\vec{q}^{\;2}}=(m_p^2\xi^2+\vec{q}^{\;2})|\Phi_B(\xi,\vec{q}^{\;2})|^2\frac{\xi}{(1-\xi)^2}\sigma_{\pi^+p}(\xi\;s),
\end{equation}
\begin{equation}
 \label{Udsigma}\frac{d\sigma(s/s_0,\xi,\vec{q}^{\;2})}{d\xi d\vec{q}^{\;2}}=S(s/s_0,\xi,\vec{q}^{\;2}) \frac{d\sigma_0(\xi,\vec{q}^{\;2})}{d\xi d\vec{q}^{\;2}},
\end{equation}
\begin{equation}
 \label{survival}S=\frac{m_p^2\xi^2 |\Phi_0(s/s_0,\xi,\vec{q}^{\;2})|^2+\vec{q}^{\;2}|\Phi_s(s/s_0,\xi,\vec{q}^{\;2})|^2}{(m_p^2\xi^2+\vec{q}^{\;2})|\Phi_B(\xi,\vec{q}^{\;2})|^2},
\end{equation}
where functions $\Phi_0$ and $\Phi_s$ arise from different spin contributions to the amplitude
\begin{equation}
\label{spinamplitude}
A_{p\to n}=\frac{1}{\sqrt{1-\xi}}\bar{\Psi}_n \left(
m_p\xi\; \hat{\sigma}_3\cdot \Phi_0+\vec{q}\;\hat{\vec{\sigma}}\cdot\Phi_s
\right) \Psi_p
\end{equation}
and both are equal to $\Phi_B$ in the Born approximation. Here $\hat{\sigma}_i$ are Pauli matrices and $\bar{\Psi}_n$, $\Psi_p$ are neutron and proton spinors. All the above functions can be calculated from the following formulae:
\begin{eqnarray}
&& \Phi_B(\xi,\vec{q}^{\;2})=\frac{N(\xi)}{2\pi}\left( 
\frac{1}{\vec{q}^{\;2}+\epsilon^2}+\imath \frac{\pi\alpha_{\pi}^{\prime}}{2(1-\xi)}
\right)\exp(-\beta^2\vec{q}^{\;2})\simeq\nonumber\\
&& \simeq \frac{N(\xi)}{2\pi}
\frac{1}{\vec{q}^{\;2}+\epsilon^2}
\frac{1}{1+\beta^2\vec{q}^{\;2}},\; \vec{q}\to 0,\\
&& N(\xi)=(1-\xi)\frac{G_{\pi^+pn}}{2}\xi^{\frac{\alpha_{\pi}^{\prime}\epsilon^2}{1-\xi}}\exp\left[ -b\frac{m_p^2\xi^2}{1-\xi}\right],\\
&&\beta^2=\frac{b+\alpha_{\pi}^{\prime}\ln\frac{1}{\xi}}{1-\xi},\; \epsilon^2=m_p^2\xi^2+m_{\pi}^2(1-\xi),
\end{eqnarray}
\begin{eqnarray}
&& \Theta_0(b,\xi,|\vec{q}|)=\frac{b\; J_0(b|\vec{q}|)\left(K_0(\epsilon\;b)-K_0\left(\frac{b}{\beta}\right)\right)}{1-\beta^2\epsilon^2},\\
&& \Theta_s(b,\xi,|\vec{q}|)=\frac{b\; J_1(b|\vec{q}|)\left(\epsilon\; K_1(\epsilon\; b)-\frac{1}{\beta}K_1\left( \frac{b}{\beta}\right)\right)}{1-\beta^2\epsilon^2},\\
&& \Phi_0=\frac{N(\xi)}{2\pi}\int\limits_0^{\infty} db\; \Theta_0(b,\xi,|\vec{q}|)V(b),\\
&& |\vec{q}|\Phi_s=\frac{N(\xi)}{2\pi}\int\limits_0^{\infty} db\;\Theta_s(b,\xi,|\vec{q}|) V(b),\\
&& V(b)=\exp\left( -\Omega_{el}(s/s_0,b)\right),\\
&& \label{U3}\Omega_{el}=\sum\limits_{i=1}^3 \Omega_i,\; \Omega_i=\frac{2c_i}{16\pi B_i}\left(\frac{s}{s_0}{\rm e}^{-\imath\frac{\pi}{2}} \right)^{\alpha_{IP_i}(0)-1}\exp\left[ -\frac{b^2}{4B_i}\right],\\
&& \label{U4}B_i=\alpha^{\prime}_{IP_i}\ln\left(\frac{s}{s_0}{\rm e}^{-\imath\frac{\pi}{2}} \right)+\frac{r_i^2}{4}, 
\end{eqnarray}
the values of parameters can be found in~(\ref{3IPtrajectories}) and in Table~\ref{tab:3IP}. 

\begin{table}
\caption{\label{tab:3IP}Parameters of the model.}
\begin{center}
\begin{tabular}{|c|c|c|c|}
\hline
 $i$        &      1   & 2   & 3  \\
\hline 
 $c_i$          &    $53.0\pm 0.8$   &   $9.68\pm 0.16$  &   $1.67\pm 0.07$ \\
\hline 
  $r^2_i$ (GeV$^{-2}$)&     $6.3096\pm 0.2522$   &    $3.1097\pm 0.1817$ &   $2.4771\pm 0.0964$\\
\hline 
\end{tabular}
\end{center}
\end{table}  
 
 For $\pi^+ p$ interaction we use, for example, the Donnachie-Landshoff parametrization~\cite{lanshofftot}
 \begin{equation}
 \label{sigland}
\sigma_{\pi^+p}(s)= 13.63\; s^{0.0808}+25.56\; s^{-0.4525},\; ({\rm mb}).
 \end{equation} 
and COMPETE~\cite{COMPETE} 
 \begin{eqnarray}
&& \label{sigcomplete}\sigma_{\pi^+p}(s)= Z_{\pi p}+B\ln^2\left(\frac{s}{s_0}\right)+\left(Y_+s^{\alpha_+}-Y_-s^{\alpha_-}\right)/s ,\; ({\rm mb}).\\
&& Z_{\pi p}=21.23\pm 0.33\; {\rm mb},\; B=0.3152\pm 0.0095\; {\rm mb},\;\\
&& s_0=34\pm 5.4\; {\rm GeV}^2,\;\\
&& Y_+=17.8\pm 1.1,\; \alpha_+=0.533\pm 0.015,\;\\
&& Y_-=5.72\pm 0.16,\; \alpha_-=0.4602\pm 0.0064.
 \end{eqnarray} 
parametrizations for $\pi^+\; p$ cross-section to make
predictions for ISR~\cite{hadr1} and PHENIX~\cite{hadr5} 

\begin{center} 
\begin{figure}[th!]
\hskip  1cm \vbox to 10cm 
{\hbox to 12cm{\epsfxsize=12cm\epsfysize=10cm\epsffile{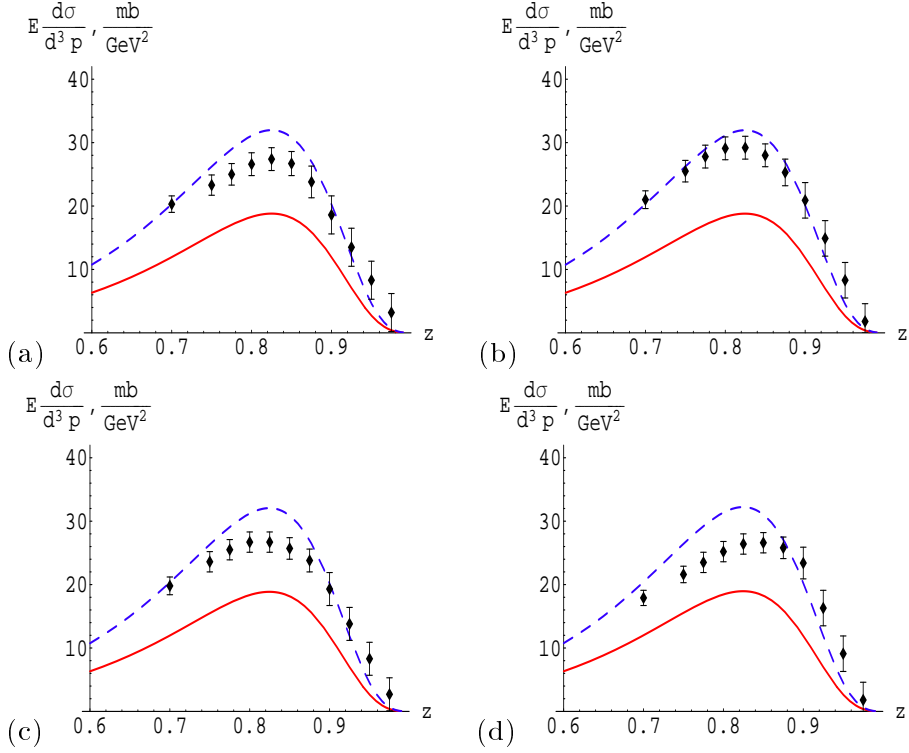}}}
\vskip 0.8cm
\caption{\label{fig:4}Theoretical differential cross-section $E\; d\sigma/d^3p$ in ${\rm mb}/{\rm GeV}^2$ versus ISR data~\cite{hadr1}: a) $\sqrt{s}=30.6$~GeV, b) $\sqrt{s}=44.9$~GeV, c) $\sqrt{s}=52.8$~GeV, d) $\sqrt{s}=62.7$~GeV. 
Lower curves are the theoretical predictions and upper curve are the predictions multiplied by factor 1.7.}
\end{figure}  
\end{center}

\begin{center} 
\begin{figure}[h!]
\hskip  1cm \vbox to 5cm 
{\hbox to 12cm{\epsfxsize=12cm\epsfysize=5cm\epsffile{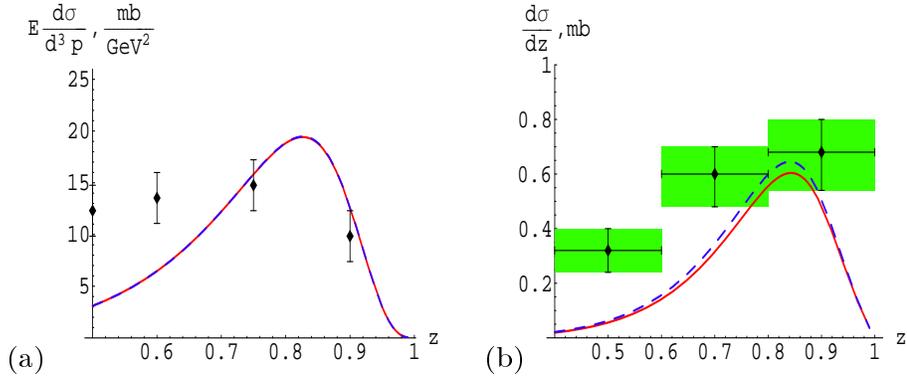}}}
\vskip 0.8cm
\caption{\label{fig:5}Theoretical differential cross-sections: a) $Ed\sigma/d^3p$ in ${\rm mb}/{\rm GeV}^2$ versus NA49 data~\cite{hadr4b} at $\sqrt{s}=17.2$~GeV; b) $d\sigma/dz$ in mb versus PHENIX data~\cite{hadr5} at $\sqrt{s}=200$~GeV, from parametrizations~(\ref{sigland}) (solid) and~(\ref{sigcomplete}) (dashed). Two curves coincides in a).}
\end{figure}  
\end{center}

The results are depicted in Fig.~\ref{fig:4} for parametrizations~(\ref{sigland}) and~(\ref{sigcomplete}). It is clear from the figures, that the calculated cross-section is about 1.7 factor lower than experimental points, this contradiction has been discussed in~\cite{workn1}. It can reflect wrong normalization of the ISR data, since the new data~\cite{hadr4b},\cite{hadr5} are lower and closer to predictions of the model as depicted in the Fig.~\ref{fig:5}. 

\begin{center} 
\begin{figure}[t!]
\hskip  3cm \vbox to 6cm 
{\hbox to 10cm{\epsfxsize=10cm\epsfysize=6cm\epsffile{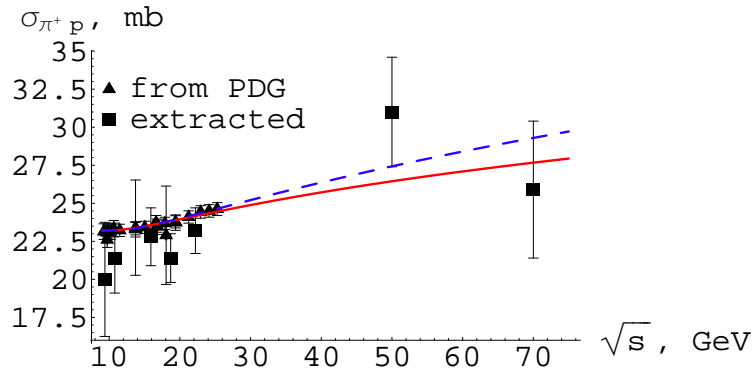}}}
\vskip 0.1cm
\caption{\label{fig:cspip} $\sigma_{\pi^+p}$ extracted from data~\cite{pipHERA1},\cite{hadr1},\cite{hadr4b},\cite{hadr5} and measured in real experiments~\cite{pipreal}. Two parametrizations~(\ref{sigland}) (solid) and~(\ref{sigcomplete}) (dashed) are also shown.}
\end{figure}  
\end{center}

We can use any other reliable parametrization. Nevertheless our principal aim is {\bf to extract $\sigma_{\pi^+p}(s)$ from the data}. For this task we need a procedure of such extraction. One of the methods~\cite{chewlow} is to evaluate the factor in front of $\sigma_{\pi^+p}$ in~(\ref{born}) at some low $t\simeq -0.014$, and then divide the data by this factor. For fixed value of $t$ we obtain $\xi\simeq 0.125$ for $\vec{q}\sim 0$. It is convenient for the estimation. We have to take the ISR data~\cite{hadr1} for
the cross-section $E\; d\sigma/d^3p$ at $x=1-\xi=0.875$ and we get an approximate formula for the total $\pi^+p$ cross-section
\begin{eqnarray}
\label{extractcs}
 \sigma_{\pi^+ p}(0.125\;s)&\simeq&\frac{\pi}{1.7\cdot (3\; {\rm GeV}^{-2})\cdot S(s/s_0,0.125,0)}\left( E\frac{d\sigma(\vec{q}\sim0)}{d^3p}\right)\simeq\nonumber\\
&\simeq&\left( 0.9\; {\rm GeV}^2\right)\left( E\frac{d\sigma(\vec{q}\sim0)}{d^3p}\right),
\end{eqnarray}
where $S(s/s_0,0.125,0)\simeq 0.68$. For low energies we obtain quite reasonable predictions for the cross-section (see Table~\ref{tab:csppiisr} and Fig.~\ref{fig:cspip}), which are compatible with real $\pi^+ p$ measurements~\cite{pipreal}, if we take into account reggeon corrections. 
\begin{table}
\caption{\label{tab:csppiisr}Total $\pi^+p$ cross-sections extracted from the ISR~\cite{hadr1}, NA49~\cite{hadr4b} (first numbers), HERA~\cite{pipHERA1} and  PHENIX~\cite{hadr5} (last two numbers) data. The last row shows numbers from COMPETE parametrization~(\ref{sigcomplete}).}
\begin{center}
{\small
\begin{tabular}{|c|c|c|c|c|c|c|c|}
\hline\label{tab:csppiisr}
 $\sqrt{s}$, GeV  & 9.4 &      10.8   & 15.9   & 18.7  &  22.2 & 50 & 70 \\
\hline 
 $\sigma_{\pi^+p}$ , mb & $20\pm 3.75$  &    $21.4\pm 2.3$   &   $22.8\pm 1.9$  &   $21.4\pm 1.6$ & $23.2\pm 1.5$ & $31\pm 3.6 $ & $25.9\pm 4.5$\\
\hline 
 $\sigma^{exp.}_{\pi^+p}$ , mb & $23.2$ & $23.19$ &  $23.55$ & $23.85$ & $24.27$ & $27.43$ & $29.3$ \\
\hline
\end{tabular}
}
\end{center}
\end{table} 

\begin{center} 
\begin{figure}[h!]
\vskip 1.2cm
\hskip  1cm \vbox to 5cm 
{\hbox to 12cm{\epsfxsize=12cm\epsfysize=5cm\epsffile{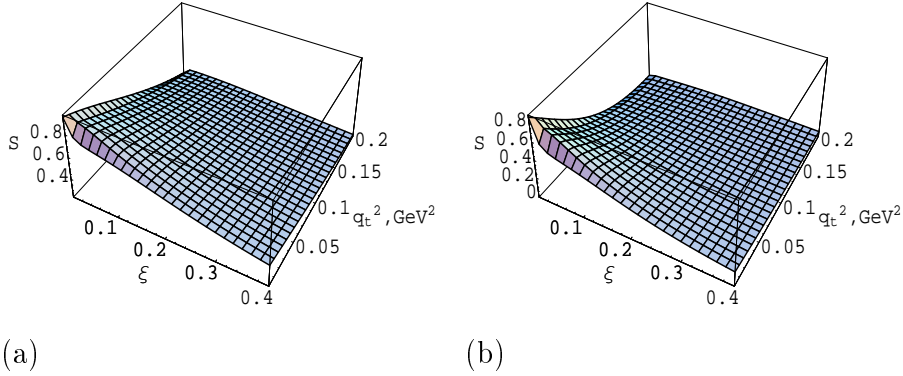}}}
\vskip 0.2cm
\caption{\label{fig:6}Function $S(s/s_0,\xi,q_t)$ at a) $\sqrt{s}=62.7$~GeV and b) $\sqrt{s}=10$~TeV.}
\end{figure}  
\end{center}

In the real situation factor $S$ is the model dependent function presented in the Fig.~\ref{fig:6}, but at $t\to 0$ it tends to unity and we can obtain more model independent cross-section $\sigma_{\pi^+p}$.
 
\begin{center} 
\begin{figure}[h!]
\hskip  1cm \vbox to 5cm 
{\hbox to 12cm{\epsfxsize=12cm\epsfysize=5cm\epsffile{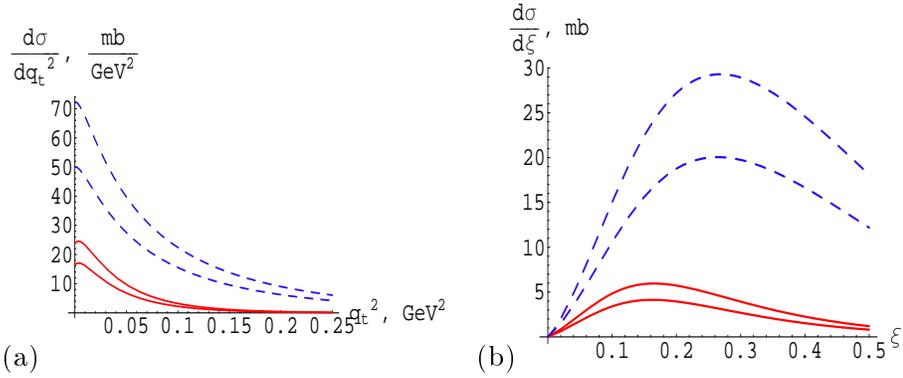}}}
\vskip 0.8cm
\caption{\label{fig:7}Cross-sections at  $\sqrt{s}=10$~TeV: a) $d\sigma/d\vec{q}^{\;2}$ and b) $d\sigma/d\xi$ for different parametrizations. Born cross-sections are depicted as dashed curves and corrected cross-sections are shown by solid curves. From each couple of curves lower curves represent Donnachie-Landshoff parametrization~(\ref{sigland})  and upper ones represent COMPETE parametrization~(\ref{sigcomplete}).}
\end{figure}  
\end{center} 

\begin{center} 
\begin{figure}[h!]
\hskip  0.1cm \vbox to 6.5cm 
{\hbox to 14cm{\epsfxsize=14cm\epsfysize=6.5cm\epsffile{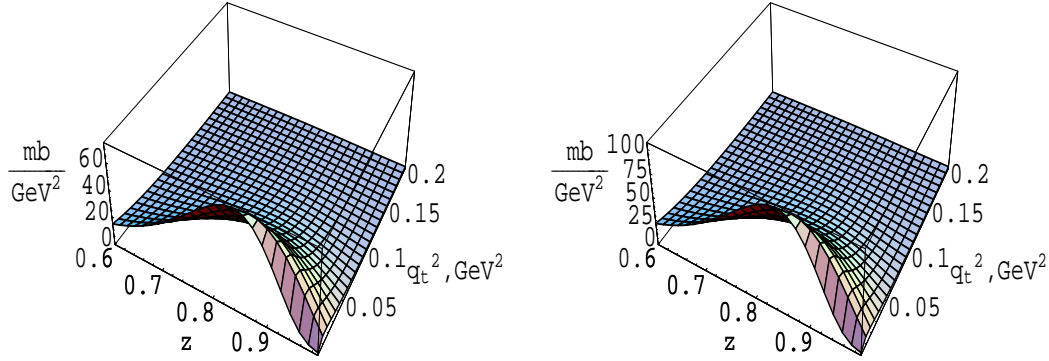}}}
\vskip 0.2cm
\caption{\label{fig:8}Cross-sections $d\sigma/d\xi d\vec{q}^{\; 2}$ at  $\sqrt{s}=10$~TeV for: a) Donnachie-Landshoff parametrization~(\ref{sigland}) and b) COMPETE one~(\ref{sigcomplete}).}
\end{figure}  
\end{center}  

 Differential cross-sections for the process $p+p\to n+X$ at $\sqrt{s}=10$~TeV are depicted in Figs.~\ref{fig:7},\ref{fig:8}. The total rate is rather large to make proposed investigations(see Table~\ref{tab:tot10TeVnum})
\begin{table}
\caption{\label{tab:tot10TeVnum}Total $p+p\to n+X$ cross-sections in the kinematical region $0<|\vec{q}|<0.5\;{\rm GeV}$,  $\xi_{min}=10^{-6}<\xi<\xi_{max}$ for two parametrizations~(\ref{sigland})((\ref{sigcomplete})).}
\begin{center}
\begin{tabular}{|c|c|c|c|c|}
\hline
 $\xi_{max}$  &      0.05   & 0.1   & 0.2 & 0.3 \\
\hline 
 $\sigma_{p+p\to n+X}$ , $\mu$b  &    $42(57)$   &   $175(244)$  &   $576(820)$  & $921(1320)$\\
\hline 
\end{tabular}
\end{center}
\end{table}  
and total absorptive corrections are about $0.35$ for $\xi_{max}<0.15$.

 For low energies the region of applicability of the presented model is usually given
 by inequalities
\begin{equation}
\label{regdif}
0.01\; {\rm GeV}^2<|t|<0.5\; {\rm GeV}^2,\; 10^{-6}<\xi<0.4,
\end{equation}
but for higher energies this region may be smaller (like $\xi<0.1$), since this corresponds
to masses $M=3$~TeV at $\sqrt{s}=10$~TeV, and for large masses it may not work.

\section{Double pion exchange}

\begin{center} 
\begin{figure}[h!]
\hskip  1cm \vbox to 8cm 
{\hbox to 12cm{\epsfxsize=12cm\epsfysize=8cm\epsffile{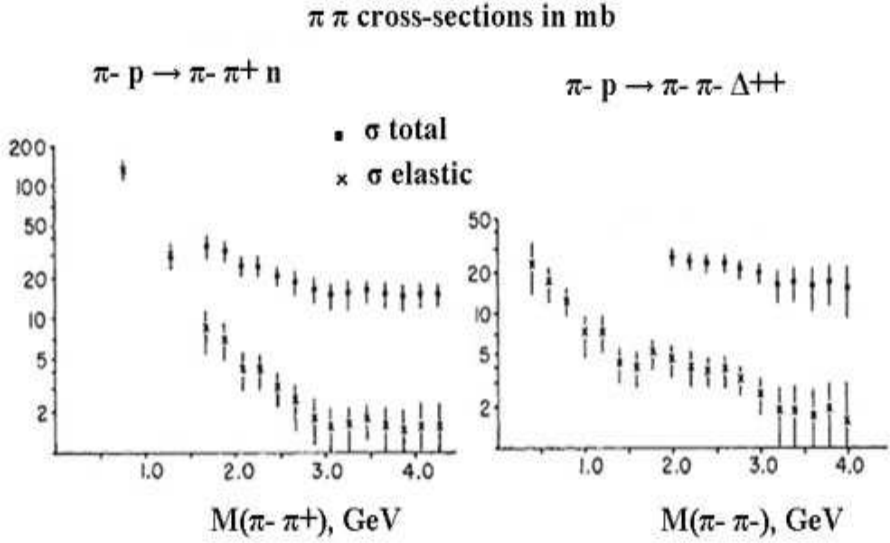}}}
\vskip 0.1cm
\caption{\label{fig:pipics} Elastic and total cross-sections for $\pi^-\pi^+$ and $\pi^-\pi^-$ scattering from the data on exclusive reactions as a function of the dipion invariant mass (Fig.5 from Ref.~\cite{pipiextract2}).}
\end{figure}
\end{center}

 As was said above the double pion exchange inclusive process is a possible source of information on both total and elastic $\pi\pi$ sross-sections. Early attempts to extract $\pi\pi$ cross-section were made with the use of the exclusive cross-section. The results are presented in Fig.~\ref{fig:pipics}~\cite{pipiextract2}. There is some tendency of the early flattening out of the $\pi\pi$ cross-sections. In $\pi p$ and $pp$ cross-sections this flattening begins at higher energies and precedes the onset of the growth. One could argue about probable early onset of the growth in $\pi\pi$ interactions.

 We can make all the above steps for the double pion exchange (Fig.~\ref{fig:2}b, D$\pi$E). Kinematics is similar to the double pomeron exchange process:
  \begin{eqnarray}
 && \label{kindpie1}p_{\pi_1}=\left( \xi_1\frac{\sqrt{s}}{2}\beta^2+\frac{t_1+m_p^2-m_n^2}{\sqrt{s}},\xi_1\frac{\sqrt{s}}{2}\beta,\vec{q}_1\right),\\
 && \label{kin1dpie2}p_{n\; 1}=\left((1-\xi_1\beta^2)\frac{\sqrt{s}}{2}-\frac{t_1+m_p^2-m_n^2}{\sqrt{s}},(1-\xi_1)\frac{\sqrt{s}}{2}\beta,-\vec{q}_1\right)\\
&& \label{kindpie3}p_{\pi_2}=\left( \xi_2\frac{\sqrt{s}}{2}\beta^2+\frac{t_2+m_p^2-m_n^2}{\sqrt{s}},-\xi_2\frac{\sqrt{s}}{2}\beta,\vec{q}_2\right),\\
 && \label{kin1dpie4}p_{n\; 2}=\left((1-\xi_2\beta^2)\frac{\sqrt{s}}{2}-\frac{t_2+m_p^2-m_n^2}{\sqrt{s}},-(1-\xi_2)\frac{\sqrt{s}}{2}\beta,-\vec{q}_2\right),
\end{eqnarray} 

\begin{eqnarray} 
 && \label{kindpie5}p_X^2=M^2=\xi_1\xi_2 s \beta^2\frac{1+\beta^2}{2}-\left(\vec{q}_1+\vec{q}_2\right)^2-m_p^2\beta^2(\xi_1^2+\xi_2^2)+\nonumber\\
 && +(t_1+t_2+2(m_p^2-m_n^2))\left(\beta^2 (\xi_1+\xi_2)+\frac{t_1+t_2+2(m_p^2-m_n^2)}{s} \right)\simeq \xi_1\xi_2 s,
 \end{eqnarray}
 \begin{equation}
  \label{kindpie6}-t_i=\frac{\vec{q}^{\;2}_i+\xi_i^2\beta^2m_p^2+(m_n^2-m_p^2)\left( \xi_i\beta^2-\frac{m_n^2-m_p^2}{s}\right)}{1-\xi_i\beta^2+\frac{2(m_n^2-m_p^2)}{s}}
 \simeq\frac{\vec{q}^{\; 2}_i+\xi_i^2m_p^2}{1-\xi_i}.
 \end{equation}
 
\begin{center} 
\begin{figure}[h!]
\vbox to 5cm 
{\hbox to 15cm{\epsfxsize=15cm\epsfysize=5cm\epsffile{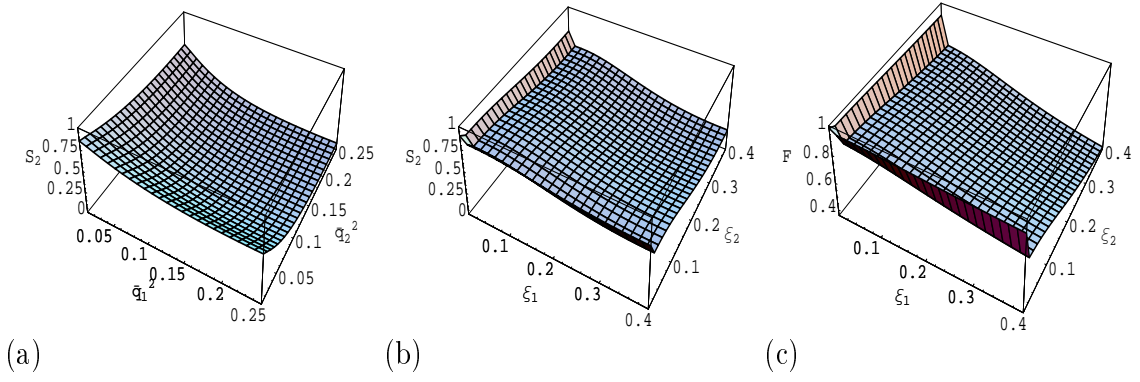}}}
\vskip 0.8cm
\caption{\label{fig:9}Function $S_2(s/s_0,\xi_{1,2},|\vec{q}_{1,2}|)$ at  $\sqrt{s}=10$~TeV for: a) fo fixed $\xi_{1,2}=0.01$ b) for fixed $|\vec{q}_{1,2}|\sim 0$. c) Function $F(\xi_1,\xi_2)$ at $\sqrt{s}=10$~TeV.}
\end{figure} 
\end{center}

Cross-section can be evaluated as follows:
\begin{eqnarray}
&& \label{dsigma0qtd}\frac{d\sigma_0(\xi_1,\xi_2,\vec{q}^{\;2}_1,\vec{q}^{\;2}_2)}{d\xi_1 d\xi_2 d\vec{q}^{\;2}_1 d\vec{q}^{\;2}_2}=\prod\limits_{i=1}^2 \left[ (m_p^2\xi_i^2+\vec{q}^{\;2}_i)|\Phi_B(\xi_i,\vec{q}^{\;2}_i)|^2 \frac{\xi_i}{(1-\xi_i)^2}\right]\sigma_{\pi^+\pi^+}(\xi_1\xi_2 s),\\
&& \label{Udsigmad}d\sigma=S_2(s/s_0,\xi_{1,2},\vec{q}^{\;2}_{1,2}) d\sigma_0,\\
&& \label{survivald}S_2=\frac{\sum\limits_{i,j=0,s} \rho_{ij}^2|\bar{\Phi}_{ij}(s/s_0,\xi_{1,2},\vec{q}^{\;2}_{1,2})|^2}{\prod\limits_{i=1}^2\left[(m_p^2\xi_i^2+\vec{q}^{\;2}_i)|\Phi_B(\xi_i,\vec{q}^{\;2}_i)|^2\right]},\\
&& \bar{\Phi}_{ij}=\frac{N(\xi_1)N(\xi_2)}{(2\pi)^2}\int\limits_0^{\infty} db_1 db_2 \Theta_i(b_1,\xi_1,|\vec{q}_1|)\Theta_j(b_2,\xi_2,|\vec{q}_2|) I_{\phi}(b_1,b_2),\\
&& I_{\phi}(b_1,b_2)=\int\limits_0^{\pi}\frac{d\phi}{\pi} V \left(\sqrt{b_1^2+b_2^2-2b_1b_2\cos\phi}\right),\\
&& \rho_{00}=m_p^2\xi_1\xi_2,\; \rho_{0s}=m_p\xi_1,\; \rho_{s0}=m_p\xi_2,\; \rho_{ss}=1.
\end{eqnarray}
For low $t_i$ function $S_2$ is approximately equal to
\begin{eqnarray}
&& F(\xi_1,\xi_2)\equiv S_2(s/s_0,\xi_1,\xi_2,0,0)\simeq\nonumber\\
&& \label{survDpiEapprox}\simeq \left( \sqrt{S(s/s_0,\xi_1,0)}+\sqrt{S(s/s_0,\xi_2,0)}- \sqrt{S(s/s_0,\xi_1,0)S(s/s_0,\xi_1,0)} \right)^2,
\end{eqnarray}
which is clear from Figs.~\ref{fig:9}b,c. Total absorptive corrections are about $0.3\div 0.5$ for $\xi_i<0.3$. Backgrounds can be estimated in the analogous way as in S$\pi$E.

 In this case we can extract $\sigma_{\pi^+\pi^+}$ from the data on D$\pi$E by formulae~(\ref{dsigma0qtd}),(\ref{Udsigmad}) at $\vec{q}_i\sim 0$.

\begin{center} 
\begin{figure}[h!]
\hskip  1cm \vbox to 12cm 
{\hbox to 12cm{\epsfxsize=12cm\epsfysize=12cm\epsffile{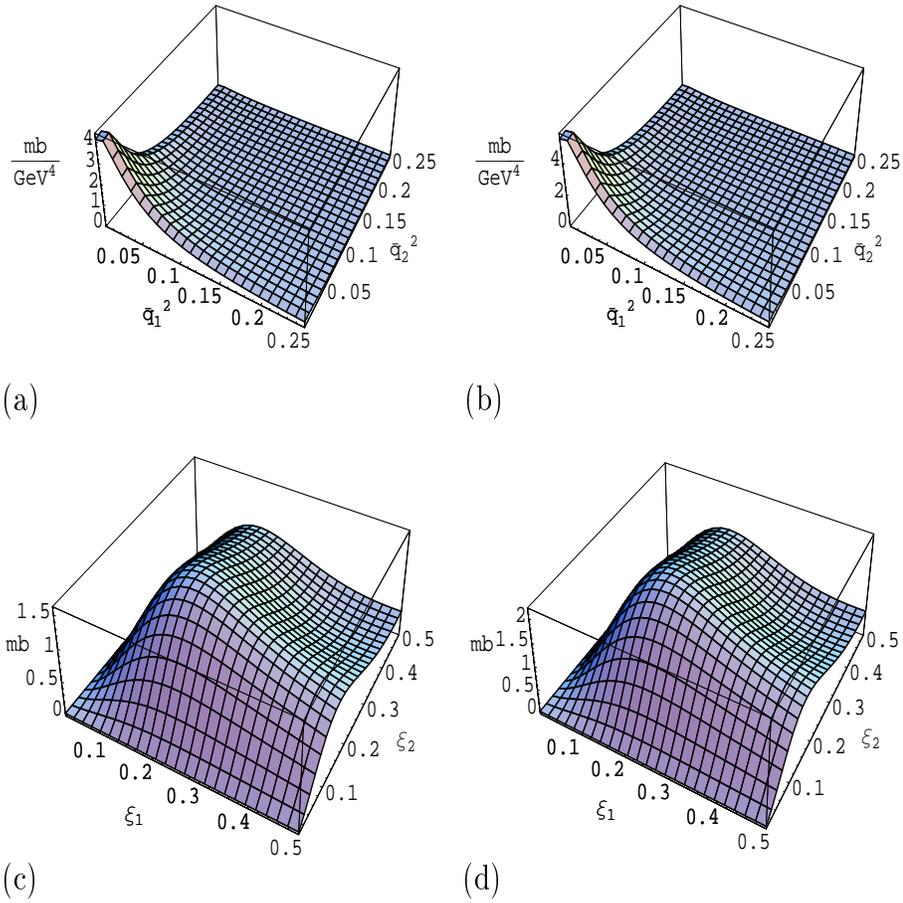}}}
\vskip 0.8cm
\caption{\label{fig:11}Integrated double cross-sections for D$\pi$E process: $d\sigma/d\vec{q}_1^{\;2}d\vec{q}_2^{\;2}$, $\xi_i<0.2$ (a,b) and $d\sigma/d\xi_1 d\xi_2$ (c,d) for (\ref{sigland}) (a,c) and (\ref{sigcomplete}) (b,d) pa\-ra\-me\-tri\-za\-tions.}
\end{figure}  
\end{center}

Since there are no data on this process, we can make only predictions for higher energies. Numerically calculated functions for D$\pi$E are presented in Figs.~\ref{fig:9},\ref{fig:11} and in Table~\ref{tab:tot10TeVDpiE} for two parametrizations.

\begin{table}
\caption{\label{tab:tot10TeVDpiE}Total $p+p\to n+X+n$ cross-sections in the kinematical region $0<|\vec{q}|<0.5\;{\rm GeV}$,  $\xi_{min}=10^{-6}<\xi<\xi_{max}$ for two parametrizations~(\ref{sigland})((\ref{sigcomplete})) multiplied by $2/3$ (quark counting rules), i.e. $\sigma_{\pi^+\pi^+}(s)=(2/3)\sigma_{\pi^+ p}(s)$.}
\begin{center}
\begin{tabular}{|c|c|c|c|c|}
\hline
 $\xi_{max}$  &      0.05   & 0.1   & 0.2 & 0.3 \\
\hline 
 $\sigma_{p+p\to n+X+n}$ , $\mu$b  &    $0.08(0.1)$   &   $1.7(2.2)$  &   $25(33)$  & $76(104)$\\
\hline 
\end{tabular}
\end{center}
\end{table}  
\vskip 1cm

\section{Experimental possibilities}
\label{section:exp}
We propose to perform measurements of S$\pi$E, 
Fig.\ref{fig:2} (a), and D$\pi$E, 
Fig.\ref{fig:2} (b),   processes at LHC with the CMS detector~\cite{CMS}.
In this chapter we discuss perspectives of such measurements at 10 TeV, 
c.m. energy of the LHC protons in the first year runs. 
For the leading neutron detection Zero Degree Calorimeter (ZDC)~\cite{ZDC} 
could be used. ZDCs are placed on the both sides of CMS at the distance 140 m from the interaction point. 
ZDC consists of electromagnetic and hadronic parts. It is designed 
for neutron and gamma measurements in the pseudorapidity region $|\eta|>8.5$. This type of detectors is widely used at RICH experiments since 2001 year~\cite{ZDCDenisov}.

\begin{figure}[ht!]
\begin{center} 
\includegraphics[width=\textwidth]{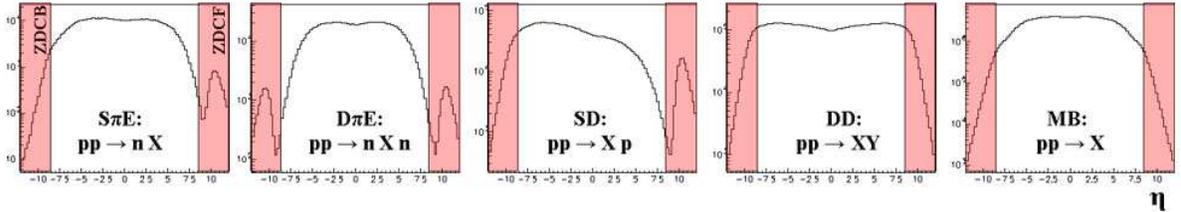} 
\caption{ Pseudorapidity distributions for signal (S$\pi$E, D$\pi$E) and background (SD, DD, MB) processes. }
\label{exp:f1}
\end{center} 
\end{figure}

To study S$\pi$E and D$\pi$E processes a generator has been developed in the framework of more 
general simulation package EDDE~\cite{EDDE}.
Kinematics of S$\pi$E and D$\pi$E reactions are defined by $\xi_n$ and $t_n$ of the 
leading neutron. Vertex $p\pi^+_{virt}n$ is generated on the basis of the 
model described above. PYTHIA 6.420 \cite{pythia} is used for the 
$\pi^+_{virt}p \to X$ generation in the S$\pi$E and $\pi^+_{virt}\pi^+_{virt} \to X$ 
generation in the D$\pi$E. 
Inelastic processes, including single and double diffractive dissociation 
(SD and DD) and minimum bias events\footnote{Usually, MB includes S$\pi$E and
D$\pi$E. Here, MB means QCD non-diffractive minimum bias with subtracted S$\pi$E and D$\pi$E
processes.} (MB), have been studied as possible 
background for S$\pi$E and D$\pi$E. All background processes have been generated 
with PYTHIA 6.420. Cross sections for signal and background at 10 TeV have 
the following ratio\footnote{Cross sections for S$\pi$E and D$\pi$E are given for 
$\xi_n<0.4$}:
$$\rm D\pi E:S\pi E:DD:SD:MB = 0.2 : 2.6 : 9.7 : 14 : 50\; mb.$$
Pseudorapidity distributions for the processes are shown in Fig.~\ref{exp:f1}. 
All processes have leading neutrons in the acceptance of ZDC, $|\eta|>8.5$. 
Thus, SD, DD, and MB can imitate S$\pi$E and D$\pi$E events and S$\pi$E can 
be a strong background for D$\pi$E measurement.

First, events have been passed through the following selections:
\begin{equation}
\label{exp:tce}
\rm
TS\pi E: 
\left[
\begin{array}{l} 
\rm N_n^f>0\quad .and.\quad N_n^b=0\quad .and.\quad \xi_n^f<0.4 \\
\rm N_n^b>0\quad .and.\quad N_n^f=0\quad .and.\quad \xi_n^b<0.4
\end{array}
\right.
\end{equation}
for S$\pi$E and
\begin{equation}
\label{exp:tdce}
\rm
TD\pi E: 
\left\{
\begin{array}{l}
\rm N_n^f>0\quad .and.\quad N_n^b>0 \\
\rm \xi_n^f<0.4\quad .and.\quad \xi_n^b<0.4
\end{array}
\right.
\end{equation}
for D$\pi$E study. Here, $N_n^f$ ($N_n^b$) is the number of neutron hitting 
the forward (backward) ZDC,
$\xi_n^f$ ($\xi_n^b$) is the relative energy loss of the forward (backward)
 leading neutron. I.e., for S$\pi$E
selection we choose events with neutrons in the forward or backward 
ZDC and with the absence of neutrons in the opposite one. For D$\pi$E, we 
select events with neutrons in the forward and 
backward ZDCs. Such selections suppress $\sim90\%$ of background events 
for S$\pi$E and suppress background for D$\pi$E by a factor of 240, 
see Tables~\ref{exp:t1} and \ref{exp:t2}. 
\begin{table}[h!]
\begin{center}
\begin{tabular}[h]{|c|ccccccccc||ccc|}
\hline
& S$\pi$E &:& D$\pi$E &:& SD &:& DD &:& MB & S &:& B \\
\hline
NOT. TS$\pi$E & 1 &:& 0.08 &:& 5.4 &:& 3.8 &:& 19.2 & 1 &:& 28.4 \\
\hline
TS$\pi$E      & 1 &:& 0    &:& 0.8 &:& 0.8 &:& 1.2 & 1 &:& 2.7 \\
\hline
\end{tabular}
\caption{Ratio of S$\pi$E and background before and after selection TS$\pi$E, 
see (\ref{exp:tce}).}
\label{exp:t1}
\end{center}
\end{table}
\begin{table}[h!]
\begin{center}
\begin{tabular}[h]{|c|ccccccccc||ccc|}
\hline
& D$\pi$E &:& S$\pi$E &:& SD &:& DD &:& MB & S &:& B \\
\hline
NOT. TD$\pi$E & 1      &:&   13   &:&    70   &:&  49   &:&  250  & 1   &:& 382 \\
\hline
TD$\pi$E      & 1      &:&   0.5  &:&    0    &:&   0.8  &:& 0.3  & 1  &:&  1.6 \\
\hline
\end{tabular}
\caption{Ratio of D$\pi$E and background before and after selection TD$\pi$E, 
see (\ref{exp:tdce}).}
\label{exp:t2}
\end{center}
\end{table}
\begin{figure}[ht!]
\begin{center} 
\includegraphics[width=\textwidth]{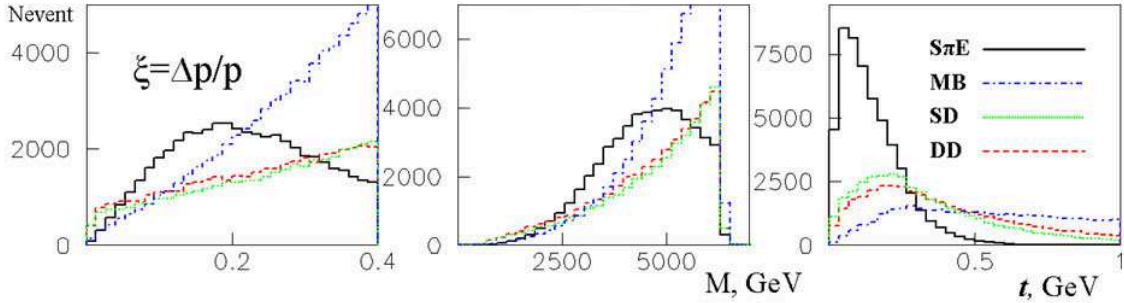} 
\caption{Distributions of $\xi$ and $t$ of the leading neutron and 
$M=\sqrt{\xi s}$
for S$\pi$E, SD, DD and MB events (selection TS$\pi$E (\ref{exp:tce})). }
\label{exp:f2}
\end{center} 
\end{figure}

\begin{figure}[ht!]
\begin{center} 
\includegraphics[width=\textwidth]{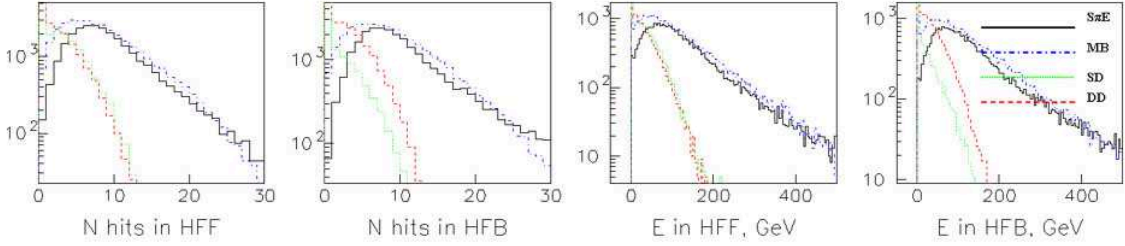} 
\caption{Number of hits and total energy deposit in the forward and backward HF.}
\label{exp:f3}
\end{center} 
\end{figure}

\noindent
Nevertheless, the signal/background ratio remains $\sim1/3$ for S$\pi$E and 
$\sim2/3$ for D$\pi$E.
Figure \ref{exp:f2} shows distributions of $\xi$ and $t$ of the leading 
neutron and $M=\sqrt{\xi s}$
after TS$\pi$E selection for S$\pi$E, SD, DD and MB events. It is seen that a cut 
in $t$ ($|t|<0.25\; {\rm GeV}^2$) should suppress 
background for S$\pi$E very efficiently. Unfortunately, in the present design of ZDC this feature is supported very restrictedly\footnote{Electromagnetic part of ZDC could be used to measure horizontal deviation of leading neutron with 50\% efficiency~\cite{Murray}}.
To search for another selections suppressing the rest of background, 
we chose S$\pi$E events with neutrons in the forward ZDC and looked at the 
data from other CMS calorimeters: Barrel, Endcap, HF, CASTOR and 
electromagnetic part of ZDC. It was found essential difference in 
the number of hits and in the total energy deposit for SD and DD 
comparing with S$\pi$E in the Barrel, Endcap and HF. As an example, we show the  
number of hits and the total energy deposit in the forward and backward 
HF for all studied processes, see Figure~\ref{exp:f3}.

\noindent
Selection 
\begin{equation}
\label{exp:thfce}
\rm
THFhitsS\pi E: 
\left[
\begin{array}{l} 
\rm N_n^f>0\quad .and.\quad N_n^b=0\quad .and.\quad \xi_n^f<0.4\quad .and.\quad N_{hits}^{HFB}>7 \\
\rm N_n^b>0\quad .and.\quad N_n^f=0\quad .and.\quad \xi_n^b<0.4\quad .and.\quad N_{hits}^{HFF}>7
\end{array}
\right.
\end{equation} 
where $\rm N_{hits}^{HF}$ is the number of hits in HF, makes SD and 
DD background negligible, but it has no any influence on MB events. 
Table \ref{exp:t3} shows signal/background ratios with selection 
(\ref{exp:thfce}). In the mass region below 5000 GeV we could expect 
S/B$\sim$10/6.

\begin{figure}[ht!]
\begin{center} 
\includegraphics[width=0.63\textwidth]{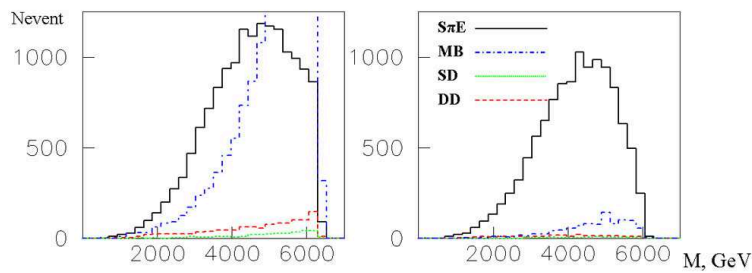} 
\caption{$M=\sqrt{\xi_n s}$ distribution for S$\pi$E and background, 
selections (\ref{exp:thfce}) (left) and (\ref{exp:ttce}) (right).} 
\label{exp:f4}
\end{center} 
\end{figure}

\noindent
For MB suppression we have to require additional cut 
using $t$ of the leading neutron:
\begin{equation}
\label{exp:ttce}
\rm
TtmaxS\pi E:\quad \left|{\it t}_n \right| < 0.2\; {\rm GeV}^2
\end{equation} 
\noindent
As shown in the table \ref{exp:t3} selection (\ref{exp:ttce}) suppresses 
MB events by a factor of 8.7 and makes S/B ratio $\sim100/8$. 
Figure \ref{exp:f4} presents $M$ distribution for S$\pi$E and background 
with selections (\ref{exp:thfce}) and (\ref{exp:ttce}). 
The same S/B ratio for D$\pi$E process could be achieved 
with selections
\begin{equation}
\label{exp:thfdce}
\rm
THFhitsD\pi E:\quad N_{hits}^{HFF}>4\quad .and.\quad N_{hits}^{HFB}>4 
\end{equation} 
and
\begin{equation}
\label{exp:ttdce}
\rm
TtmaxD\pi E:\quad \left|{\it t}_n^f \right| < 0.3\; {\rm GeV}^2\quad .and.\quad \left|{\it t}_n^b \right| < 0.3\; {\rm GeV}^2.
\end{equation} 
 
\begin{table}[ht!]
\begin{center}
\begin{tabular}[h]{|c|ccccccc||ccc|}
\hline
& S$\pi$E &:& SD &:& DD &:& MB & S &:& B \\
\hline
TNhitsS$\pi$E              & 1 &:& 0.013 &:& 0.06 &:& 0.48  & 1 &:& 0.56 \\
\hline
TNhitsS$\pi$E.and.TRtmaxS$\pi$E & 1 &:& 0.005 &:& 0.02 &:& 0.055 & 1 &:& 0.08 \\
\hline
\end{tabular}
\caption{Ratio of S$\pi$E to background with selections TNhitsS$\pi$E 
(\ref{exp:thfce}) and TtmaxS$\pi$E (\ref{exp:ttce}).}
\label{exp:t3}
\end{center}
\end{table}
\begin{table}[h!]
\begin{center}
\begin{tabular}[h]{|c|ccccccccc||ccc|}
\hline
& D$\pi$E &:& S$\pi$E &:& SD &:& DD &:& MB & S &:& B \\
\hline
TNhitsD$\pi$E                 & 1  &:&  0.47  &:&  0.0   &:&   0.03  &:&  0.2   & 1   &:& 0.7 \\
\hline
TNhitsD$\pi$E.and.TRtmaxD$\pi$E   & 1  &:&  0.06  &:&  0.0   &:&  0.005  &:&  0.004  & 1  &:&  0.07 \\
\hline
\end{tabular}
\caption{Ratio of D$\pi$E to background with selections 
TNhitsD$\pi$E~(\ref{exp:thfdce}) and TtmaxD$\pi$E~(\ref{exp:ttdce}).}
\label{exp:t4}
\end{center}
\end{table}
\begin{figure}[ht!]
\begin{center} 
\includegraphics[width=0.63\textwidth]{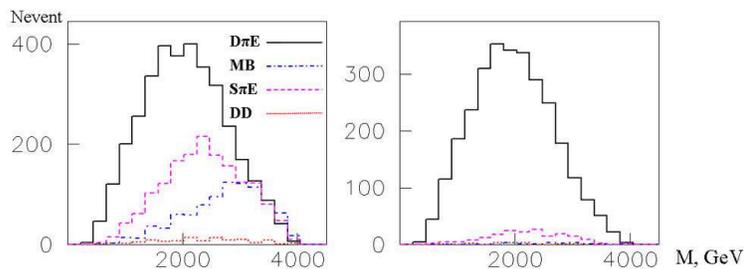} 
\caption{Distribution in $M=\sqrt{\xi_n^f\xi_n^b s}$ for D$\pi$E and background, 
selections (\ref{exp:thfdce}) (left) and (\ref{exp:ttdce}) (right).} 
\label{exp:f5}
\end{center} 
\end{figure}

\noindent
Table \ref{exp:t4} shows ratios for D$\pi$E and background with 
selections (\ref{exp:thfdce}) and (\ref{exp:ttdce}). 
Figure \ref{exp:f4} presents $M$ distribution for D$\pi$E and background 
with selections (\ref{exp:thfdce}) and (\ref{exp:ttdce}). 

\begin{center} 
\begin{figure}[t!]
\hskip  1cm \vbox to 6cm 
{\hbox to 12cm{\epsfxsize=12cm\epsfysize=6cm\epsffile{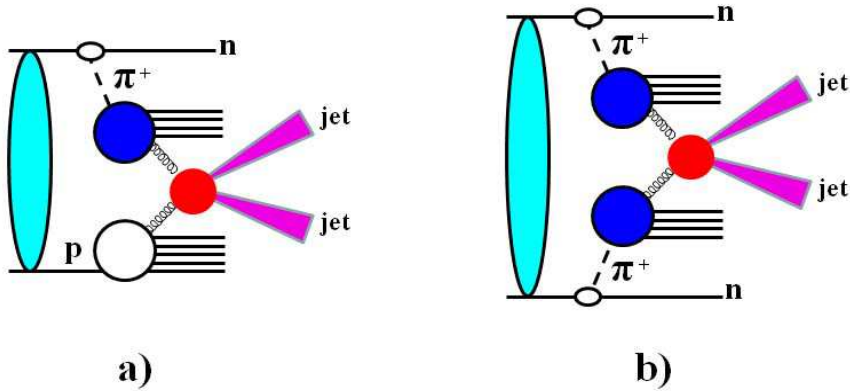}}}
\vskip -0.4cm
\caption{\label{fig:distr} Processes for the measurements of parton distributions in pions.}
\end{figure}  
\end{center}  
\section{Discussions and conclusions}

In conclusion, our study of S$\pi$E and D$\pi$E processes and background on 
the generator level
shows that we could expect S$\pi$E and D$\pi$E observation at LHC with CMS in the 
first runs as realistic. Some modifications of ZDC is required to measure $t$ of the 
leading neutron. Using combination of 2 simple cuts, for N hits in HF 
and $t_n$, one could suppress background by a factor of $\sim760$ for 
S$\pi$E and $\sim9500$ for D$\pi$E, saving 
$\sim35 \%$ of S$\pi$E events and $\sim60 \%$ of D$\pi$E. Without ZDC modification we
could expect observation of S$\pi$E events mixing with MB events in proportions 
S$\pi$E/MB $\sim10/6$ and
D$\pi$E mixing with S$\pi$E and MB in proportions D$\pi$E/(MB+S$\pi$E) $\sim10/(5+2)$. 
Thus, we could estimate
cross sections of S$\pi$E and D$\pi$E production at 10 TeV. 
For more realistic estimations a full MC study should be performed including
 detector simulation and pile-up background.

 As was said the main motivation for this work is the extraction of the total $\pi p$ and $\pi\pi$ cross-sections from proton-proton scattering measurements. Procedure is quite simple for low values of t, because absorptive factor goes to unity and backgrounds are completely suppressed. If this task is done than we will have additional, more rich, data in the high energy region to check predictions of different models for strong interactions, quark counting rules and so on. 
 
  Further task is the exact estimation of backgrounds presented in Figs.~\ref{fig:1}b,c,d (especially $\rho$, $a_2$ exchanges) to be sure that the extraction procedure is correct.
  
  To make some estimations we use several parametrizations for the total cross-sections. All they show similar behaviour. That is why we can use them to make Monte-Carlo simulations to see possible experimental difficulties. Unfortunately, the present data on S$\pi$E is not so clear (problems in normalization), so we can make only more or less plausible estimations for high energies (especially for the LHC).
  
  At the Monte-Carlo simulation on the generator level we can see that the main background can arise from minimum bias events (if we integrate cross-sections in variable t). We have found that signal to the minimum bias background ratio for S$\pi$E and D$\pi$E is about 1.5-2. It can be increased significantly only if we take a cut $|t|<0.2\div 0.3\; {\rm GeV}^2$ (which was done, for example, in low-energy experiments). Since that the  precision measurement of $t_n$ is not possible with the present design of ZDC, extraction of $\pi p$ and $\pi\pi$ cross-sections sets aside for the future ZDC modifications\footnote{Such modifications could be made later~\cite{Murray}}. Nevertheless, we can look for S$\pi$E and D$\pi$E events in the fist LHC run for rough estimations of cross-sections.

We have also some perspectives of measurements of the parton distribution functions in pions in hard processes (see Fig.~\ref{fig:distr}). This would be also very interesting, since we can get parton distributions in the pion at smaller $x$ and higher $Q^2$. But cross-sections for hard processes are small and this task is also for high luminocity runs. It seems to be fairly realistic, especially taking into account new prospects for high-energy $\pi p$ interactions. Investigation of S$\pi$E and D$\pi$E processes can also provide us with unique measurements of $\pi p$ and $\pi\pi$ elastic cross-sections.

In spite of all difficulties, proposed measurements are so exciting that it makes all sense to pursue our aim further. 
  
\section*{Acknowledgements}

We are grateful to M.~Albrow, M.~Arneodo, V.I.~Kryshkin, M.~Murray, K.~Piotrzkowski and N.E.~Tyurin for useful duscussions and helpfull suggestions. V.A.P. thanks M.G.~Ryskin for stimulating comments on inclusive neutron production.

\end{document}